\documentclass{pasj00}

\begin{document}
\SetRunningHead{Yoshitaka Hanabata et al.}{X-Ray Observations of W51 with Suzaku}
\Received{2012/08/03}
\Accepted{2012/11/27}

\title{X-Ray Observations of the W51 Complex with Suzaku}

\author{
Yoshitaka \textsc{Hanabata}\altaffilmark{1}
Makoto \textsc{Sawada}\altaffilmark{2,3}
Hideaki \textsc{Katagiri}\altaffilmark{4}
Aya \textsc{Bamba}\altaffilmark{2}
and
Yasushi {\sc Fukazawa}\altaffilmark{1}}
\altaffiltext{1}{Department of Physical Sciences, Hiroshima University, 1-3-1, Kagamiyama, Higashi-Hiroshima, Hiroshima 739-8526}
\email{hanabata@hep01.hepl.hiroshima-u.ac.jp}
\altaffiltext{2}{Department of Physics and Mathematics, Aoyama Gakuin University, 5-10-1 Fuchinobe, Sagamihara, Kanagawa 252-5258}
\altaffiltext{3}{Department of Physics, Graduate School of Science, Kyoto University, Kitashirakawa Omiwake-cho, Sakyo-ku, Kyoto 606-8502}
\altaffiltext{4}{College of Science, Ibaraki University, 2-1-1 Bunkyo, Mito, Ibaraki 310-8512}

\maketitle

\begin{abstract}
We present a detailed analysis of the X-ray emission from the middle-aged supernova remnant W51C and star-forming region W51B with Suzaku.
The soft X-ray emission from W51C is well represented by an optically thin thermal plasma in the non-equilibrium ionization state with a temperature of $\sim$0.7~keV. The elemental abundance of Mg is significantly higher than the solar value.
We find no significant feature of an over-ionized plasma in W51C. 
The hard X-ray emission is spatially coincident with the molecular clouds associated with W51B, overlapping with W51C.
 The spectrum is represented by an optically thin thermal plasma with a temperature of $\sim$5~keV or a powerlaw model with a photon index of $\sim$2.2.
 The emission probably has diffuse nature since its luminosity of 1$\times$10$^{34}$~erg~s$^{-1}$ in the 0.5--10~keV band cannot be explained by the emission from point sources in  this region.  
We discuss the possibility that the hard X-ray emission comes from stellar winds of OB stars in W51B or accelerated particles in W51C.

\end{abstract}

\section{Introduction}

Galactic cosmic rays (mostly protons) are widely believed to be accelerated at the shock
of supernova remnants (SNRs) through the diffusive shock acceleration
mechanism~\citep[and references therein]{Reynolds2008}.
TeV $\gamma$-ray observations of SNRs gave us firm evidences of acceleration of TeV particles (e.g.,~\cite{Aharonian04}).
In addition, the signatures of proton acceleration were obtained by GeV $\gamma$-ray observations of middle-aged SNRs interacting with molecular clouds (e.g.,~\cite{W51C}).
On the other hand, how the acceleration process evolves in SNRs is currently not well understood although it is crucial in understanding the acceleration mechanism in SNRs.

Theoretically, one possible scenario of the evolution of acceleration in SNRs is that efficient acceleration, accelerating a large amount of particles up to high energy, occurs at the free expansion phase and it ends at the early Sedov phase.
The cosmic-ray streaming instability becomes less efficient as the shock velocity decreases with time, and nonlinear wave interactions reduce the level of turbulence at the late Sedov stage~\citep{Volk88}. An alternative scenario is that moderate acceleration lasts until the end of the Sedov phase. In this case, tens-of-TeV particles can exist in middle-aged SNRs~\citep{Sturner97,Nakamura12}.
Thus it is important to investigate nonthermal particles in middle-aged SNRs.

The spectrum of high energy electrons traced by synchrotron X-rays could have an important clue of the evolution of the acceleration mechanism since its cooling time is fairly shorter than that of protons and the electron spectrum can be strongly affected by the cooling process.
Possible synchrotron X-ray emission from a middle-aged SNR was detected in W44~\citep{Uchida12}.
However, the number of the samples is limited because emission from the thermal plasma of middle-aged SNRs dominates in the X-ray band and nonthermal emission is buried in it.
Thus, accurate measurement of the thermal component is required to detect the nonthermal emission, in addition to observations with a highly sensitive detector in the hard X-ray band.

Moreover, investigations of SNR plasmas can also constrain the time evolution of particle acceleration.
Suprathermal particles injected into the cosmic-ray acceleration can effectively ionize ions in plasma and generate an over-ionized plasma which has a higher ionization temperature ($T_{\rm z}$) than the electron one ($T_{\rm e}$:~\cite{Kato92}).
In big solar flares, the ionization rate of plasma increases as the flux of hard X-rays from bremsstrahlung emission increases~\citep{Tanaka86}.
The degree of over-ionization depends on the elapsed time of the ionization and the density of suprathermal electrons.
Thus, we can constrain the density of accelerated cosmic rays and the time when the acceleration occurred in the SNR via the diagnosis of thermal plasma with X-ray.

W51C (G49.2$-$0.7) is an SNR with the age of $\sim$3$\times$10$^{4}$~yr  derived from a Sedov analysis with the observed X-ray temperature and the angular radius~\citep{Koo95}.
The extended $\gamma$-ray emission has been detected by the Fermi satellite, the H.E.S.S. and the MAGIC telescopes~\citep{W51C,HESS,MAGIC}. 
The $\gamma$ rays are naturally explained by $\pi^{0}$-decay through the interaction of protons accelerated in the SNR with dense gas~\citep{W51C,MAGIC}.
Thus, W51C is a good SNR to study cosmic-ray acceleration during the Sedov phase.
This SNR is located at the tangential point of the Sagittarius Arm where the distance is estimated to be 6~kpc~\citep{Koo95}. 
The remnant appears as an incomplete shell of $\sim$30$'$ diameter in the radio continuum~\citep{Subrahmanyan95}. 
The western part of the remnant is superposed upon the massive star-forming region W51B~\citep{Koo99,Nanda04}.
\citet{Koo97a} found that W51C interacts with high-velocity molecular stream and 
\citet{Carpenter98} revealed that it most likely belongs to the cloud associated with H\emissiontype{II} regions of W51B by the CO observation.
The detections of two OH (1720 MHz) maser spots give us another evidence of interaction between W51C and a part of molecular cloud associated with W51B~\citep{Green97}.

X-ray observations of W51C were carried out by Einstein, ROSAT,
ASCA and Chandra~\citep{Seward90,Koo95,Koo02,Koo05}. The soft X-ray
emission from the SNR was represented by the plasma emission in collisional
ionization equilibrium (CIE) with the temperature of 0.3--0.5~keV.
No hint of the over-ionized plasma has been reported from W51C.
In the hard X-ray band, ASCA and Chandra detected two
pulsar wind nebula (PWN) candidates and some H\emissiontype{II} regions
in addition to the extended emission overlapped with molecular clouds
of W51B~\citep{Koo02,Koo05}.
However, because of the low sensitivity of ASCA and Chandra for hard X-rays, the spectral analysis for the extended emission was not performed and its origin is unclear.

The aim of this paper is to search for the over-ionized plasma and reveal the origin of the extended hard X-rays for studying the evolution of acceleration process in middle-aged SNRs.
This study was done with the X-ray Imaging Spectrometer (XIS:~\cite{Koyama07}) onboard Suzaku~\citep{Mitsuda07} with the superior energy resolution and the large effective area even in the hard X-ray band.

\section{Observation and Data Reduction}
\label{obs}

\begin{figure}
  \begin{center}
    \FigureFile(80mm,30mm){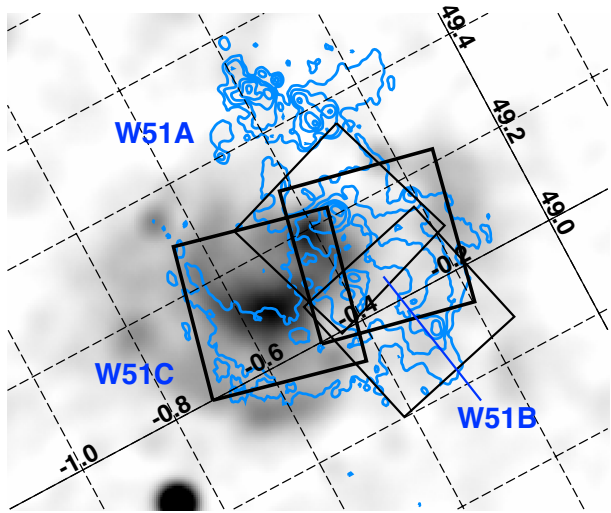}
  \end{center}
  \caption{The ROSAT image (gray-scale; 0.7--2.0~keV) of W51C region~\citep{Koo95}. 
  The thick and thin solid boxes indicate the position of XIS and Chandra field of views, respectively.
  The blue contours show the VLA 1.4~GHz radio image~\citep{Koo97a} with a level of 0.015, 0.06, 0.2, 0.6 and 1.5~Jy~beam$^{-1}$.}
\label{fig:cmap_wide}
\end{figure}

We carried out two observations around the W51C region.
Locations of two pointing fields and an observation log are shown in
figure~\ref{fig:cmap_wide} and table~\ref{tab:obs_log}, respectively.

The XIS consists of four X-ray CCD cameras, each mounted on the focal plane of the X-ray Telescope~(XRT:~\cite{Serlemitsos07}). 
Three sensors employ Front-Illuminated
(FI) CCDs (XIS~0, 2 and 3) while the other employs a Back-Illuminated
(BI) CCD (XIS~1). The entire region of XIS~2 and an edge of XIS~0 have
not been functional due to the anomalies in 2006 November and in 2009 June,
respectively. 
The XRT has a point spread function of $\sim$2$'$. 
The total effective areas is about 400~cm$^{2}$ at 8~keV.

The XIS was operated in the normal clocking full-window mode, and the
data of the two editing modes, 3$\times$3 and 5$\times$5, were
combined. 
To restore the radiation-induced degradation in the energy gain and resolution, the spaced-row charge injection technique~\citep{Bautz04} was applied with the makepi files version 20110708 provided by the XIS team~\citep{Uchiyama09}.
In the analysis, we employed the data with a process version 2.5 for W51C\_W and W51C\_E and a version 2.4 for the background data, respectively.
The data were analyzed with the HEAsoft version 6.11.
We removed hot and flickering pixels, and discarded events with a condition during the South Atlantic Anomaly passages, Earth night-time elevation angle below 5$^{\circ}$, and day-time angle below 20$^{\circ}$.

We also used Chandra archival data of observations around W51C to evaluate the contribution of point sources to the extended hard X-rays. The observation log is also shown in table~\ref{tab:obs_log}.

\begin{table*}[hbtp]
  \begin{center}
  \caption{Log of Suzaku and Chandra observations of W51C region.}\label{tab:obs_log}
    \begin{tabular}{cccccc}\hline
     Instrument & Field name & Sequence no. & Aim point\footnotemark[$*$] & Start date & Effective \\
     &   &   & ($l$,$b$) & (UTC)  & exposure\\
     \hline
     Suzaku & W51C\_W   & 504066010 & ($49\fdg11, -0\fdg31$) & 2010/03/28 &
		     44.1~ks \\
     & W51C\_E   & 504067010 & ($49\fdg11, -0\fdg54$) & 2010/03/30 &
		     43.7~ks  \\
      & BGD & 504044010 & ($65\fdg87, -0\fdg32$) & 2009/05/23 &
		     38.7~ks  \\   \hline
    Chandra & W51B & 500318 & ($48\fdg96, -0\fdg31$) & 2003/06/03 &
		     29.9~ks \\
     & W51\_NORTH & 500319 &   ($49\fdg17, -0\fdg35$) & 2002/12/08 &
		     11.8~ks \\
     \hline
     \multicolumn{6}{l}{\footnotemark[$*$] In Galactic coordinates.}\\
    \end{tabular}
  \end{center}
\end{table*}

\section{Analysis}
\label{ana}

\subsection{Imaging Analysis}
\label{image_ana}

\begin{figure*}
  \begin{center}
    \FigureFile(80mm,30mm){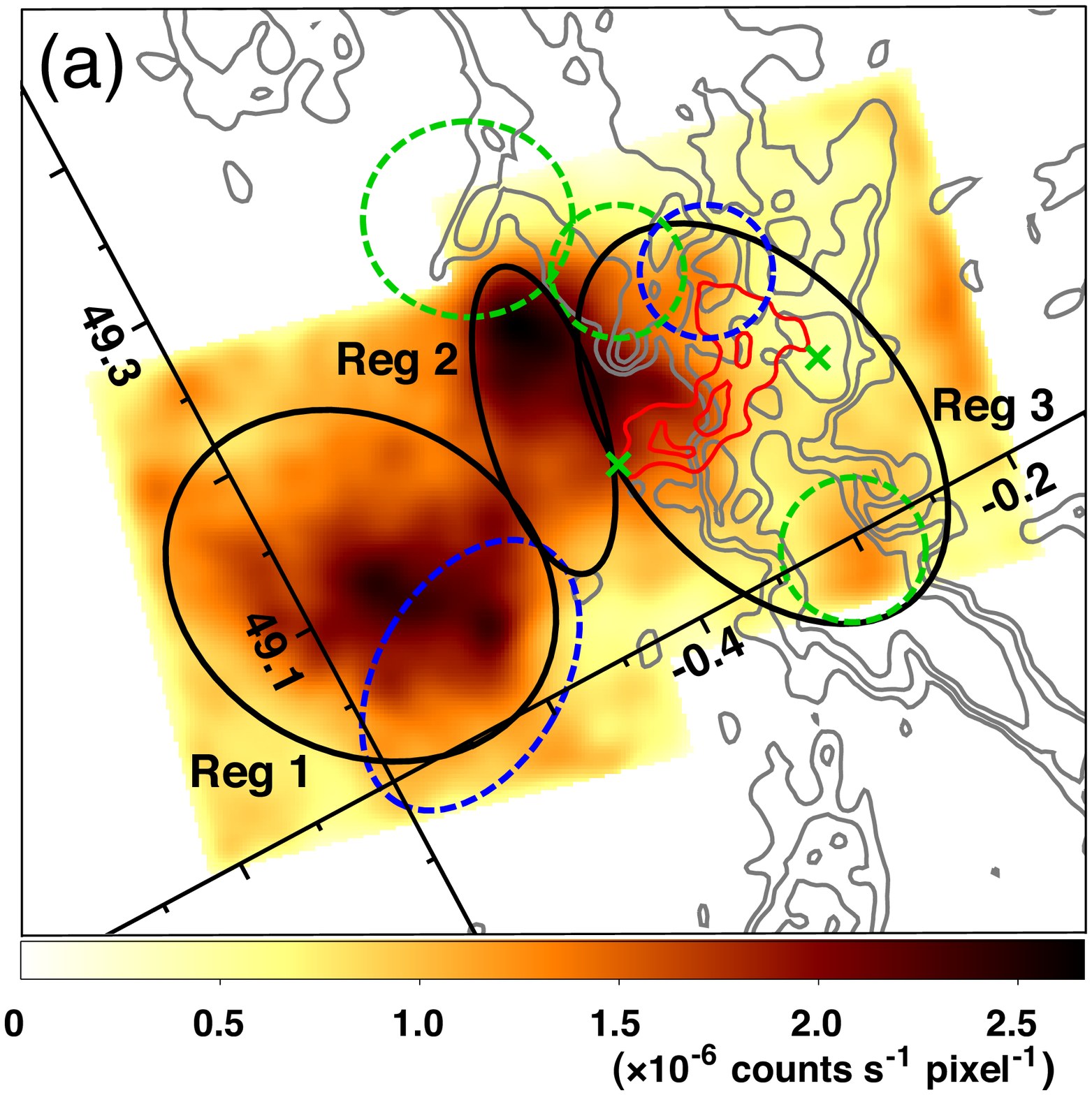}
    \FigureFile(80mm,30mm){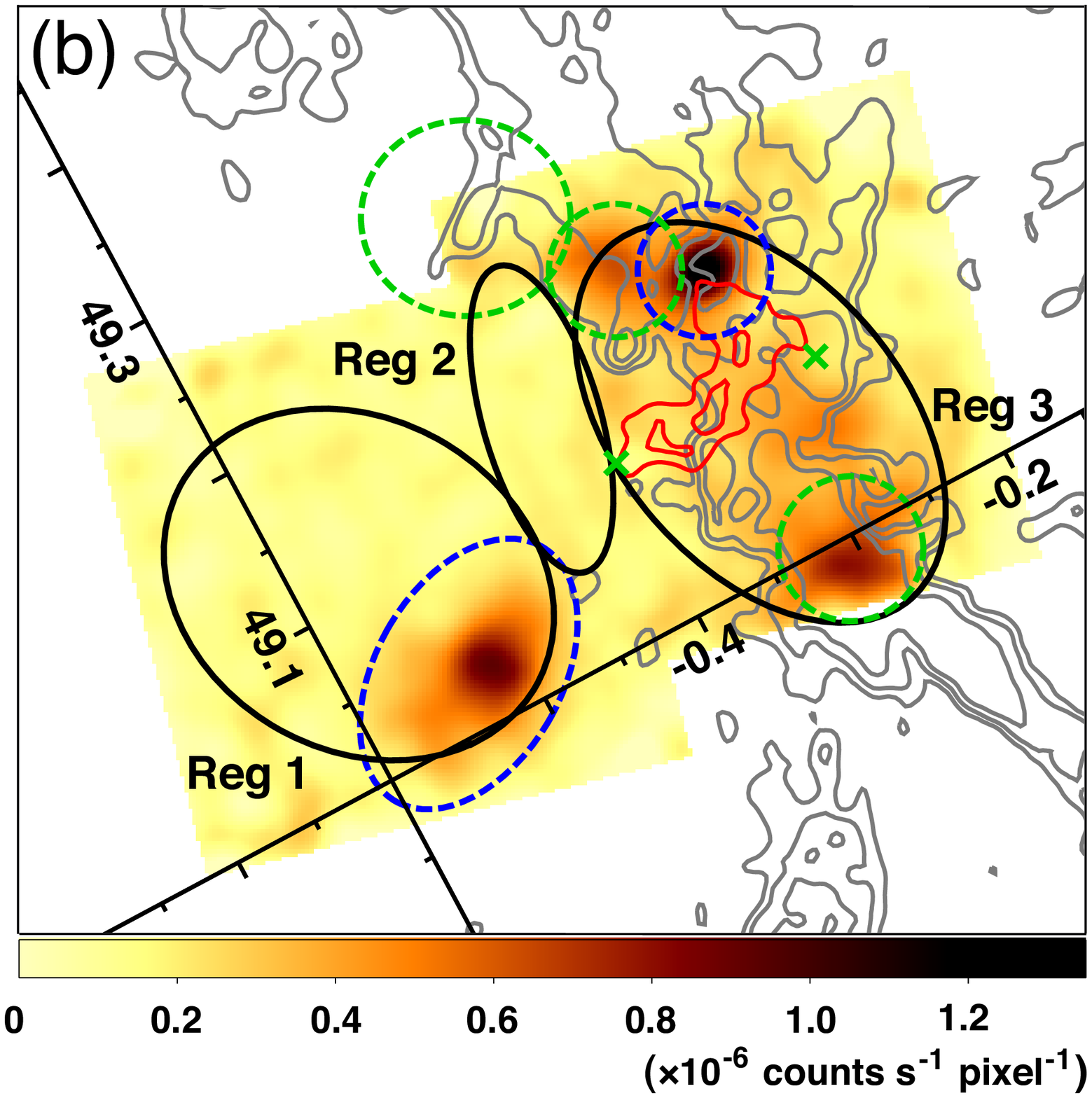}
  \end{center}
  \caption{Suzaku XIS counts map of W51C region in the (a) 0.5--2.5~keV
 and (b) 3.5--8.0~keV bands.
Gray contours show $^{12}$CO (J~=~2$-$1) line
 intensity~\citep{Bieging10} at 10, 20 and 40\% levels, for the velocity
 range from 60 to 70~km~s$^{-1}$. Red contours indicate the intensity distribution of the high
 velocity H\emissiontype{I} gas at 40 and 80\% levels, integrated in the velocity
 range from 85 to 144~km~s$^{-1}$. Black ellipses indicate the regions
 used for the spectral analyses. 
Dashed regions indicate the excluded regions for the analysis, where two blue ones, a north large green one, and the other green ones are PWN candidates, calibration source, and H\emissiontype{II} region of G49.0$-$0.3 and G49.2$-$0.3, respectively. 
 Two green crosses indicate the H\emissiontype{II} region of G49.1$-$0.4 and G49.10$-$0.27.}
\label{fig:cmap}
\end{figure*}

Figure~\ref{fig:cmap} shows the XIS mosaic images around W51C in the energy
bands of 0.5--2.5~keV and 3.5--8.0~keV using XIS~1 and 3.
The non-X-ray background (NXB) created by {\tt xisnxbgen}~\citep{Tawa08} was subtracted and then vignetting and exposure corrections were performed.

The W51C\_E observation covers the central bright region of the X-ray emission
from W51C, while the W51C\_W covers the
large portion of the star-forming region W51B (see figure~\ref{fig:cmap_wide}).
The 0.5--2.5~keV image shows an extended emission along the east-center
direction in the field, which has two bright portions located at southeast and central north (figure~\ref{fig:cmap}a; Reg~1 and Reg~2). These structures are consistent with the images obtained by ROSAT~\citep{Koo95} and ASCA 0.7--2.5~keV band~\citep{Koo02}.

W51C interacts with some part of the molecular cloud to which W51B most likely belongs~\citep{Koo97a,Green97}.
To investigate the correlation between the
X-rays and the molecular clouds, 
we overlaid the $^{12}$CO(J~=~2--1) line intensity map~\citep{Bieging10} for the
velocity range from 60 to 70~km~s$^{-1}$ on the XIS images. 
In the soft X-ray image, the surface brightness drops down toward the western edge of the clouds. \citet{Koo95} and \citet{Koo02} concluded that it was caused by the absorption of the intervening clouds.
The 3.5--8.0~keV image shows an extended ($\sim$8$'\times$12$'$) emission, which overlaps with the molecular clouds at the western part of the SNR (figure~\ref{fig:cmap}b; Reg~3).
There are also two PWN candidates and H\emissiontype{II} regions G49.0$-$0.3 and G49.2$-$0.3 in the hard X-ray image. In this paper, we focus on the emission from W51C and the extended hard X-rays.

We also compared the X-ray emission with the shocked H\emissiontype{I} clouds with a velocity range from 85 to 144~km~s$^{-1}$~\citep{Koo97a}. 
The clouds are located near the western boundary of the soft X-rays, while the correlation between the hard X-ray emission and the clouds is not clear.

We first investigate the spectra of bright regions in the soft X-ray band (subsection~\ref{soft_spec}); Reg~1 and Reg~2, which could be the parts of the SNR. Then we inspect the spectrum of the hard X-ray emission in Reg~3 (subsection \ref{hard_spec}).

In the spectral analyses, we used the SPEX software~\citep{Kaastra96} version 2.02.04. 
The redistribution matrix and the auxiliary response functions were generated by {\tt xisrmfgen} and {\tt xissimarfgen}~\citep{Ishisaki07}, respectively, with the rmfparam files released on 20110804.
The errors are quoted at the 90\% confidence level unless otherwise mentioned.

\subsection{Spectral Analysis of Soft X-Ray Emission}
\label{soft_spec}

\subsubsection{Background Estimation}
\label{soft_bgd}

\begin{figure*}
  \begin{center}
    \FigureFile(80mm,30mm){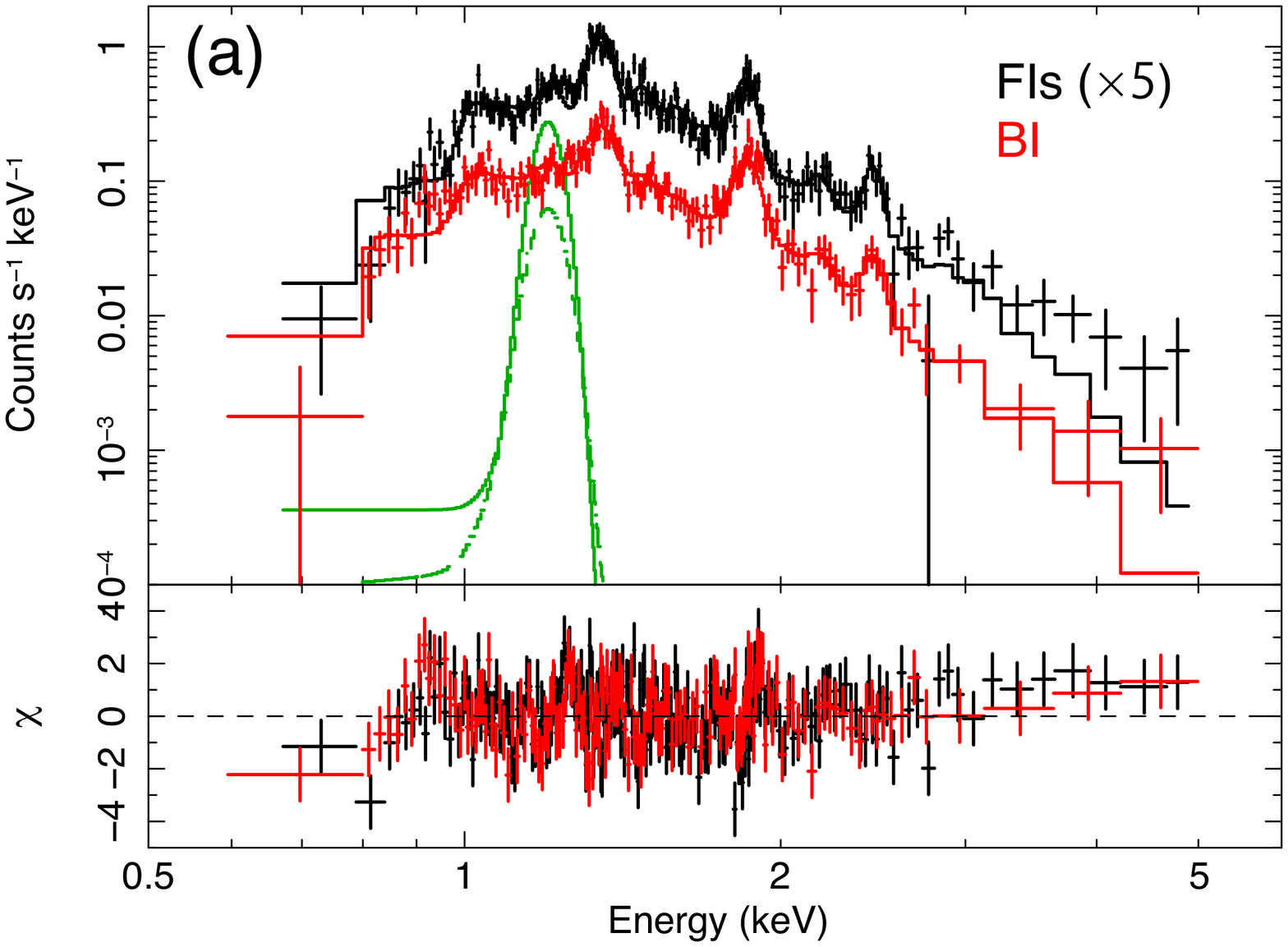}
    \FigureFile(80mm,30mm){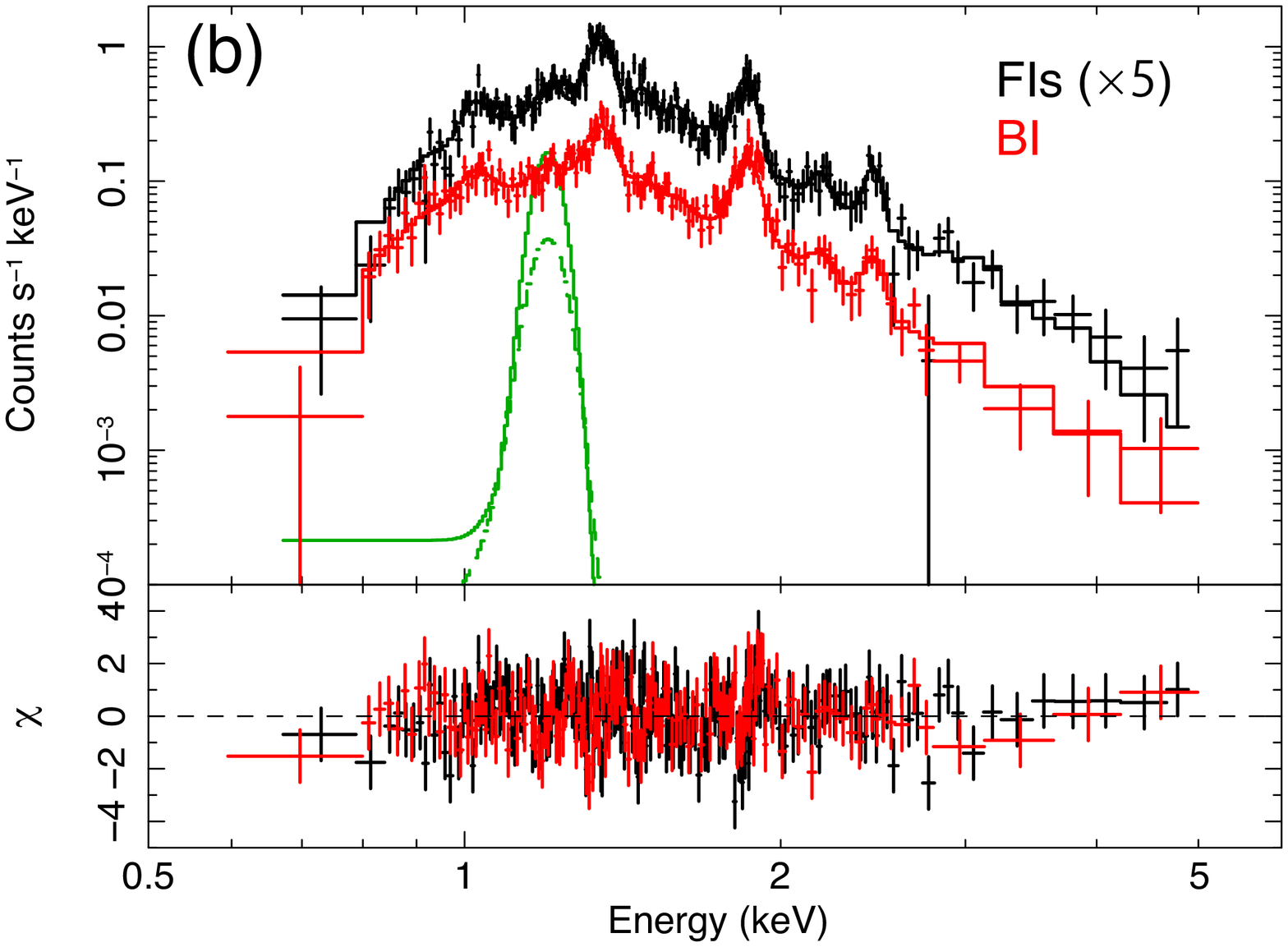}
    \FigureFile(80mm,30mm){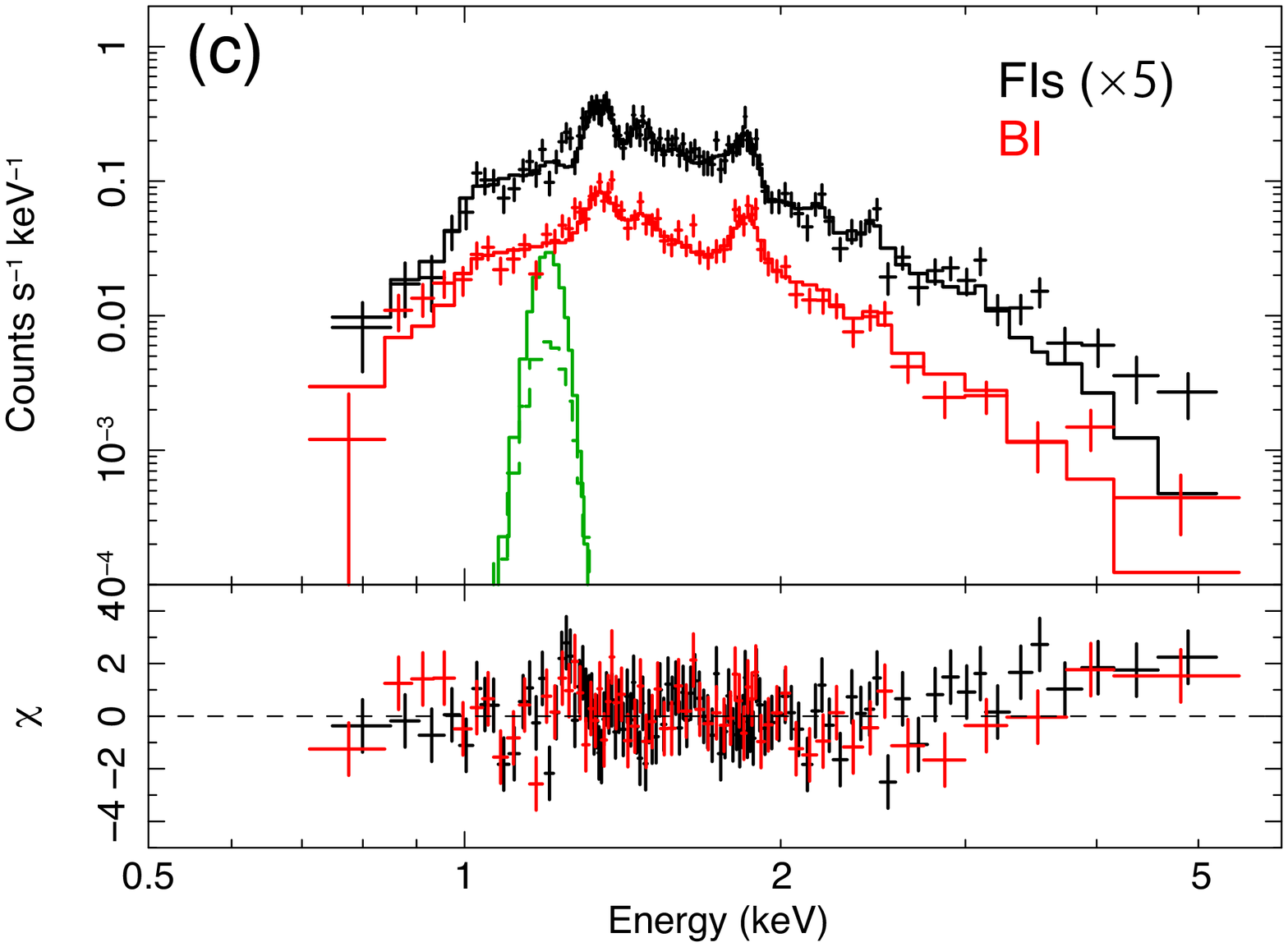}
    \FigureFile(80mm,30mm){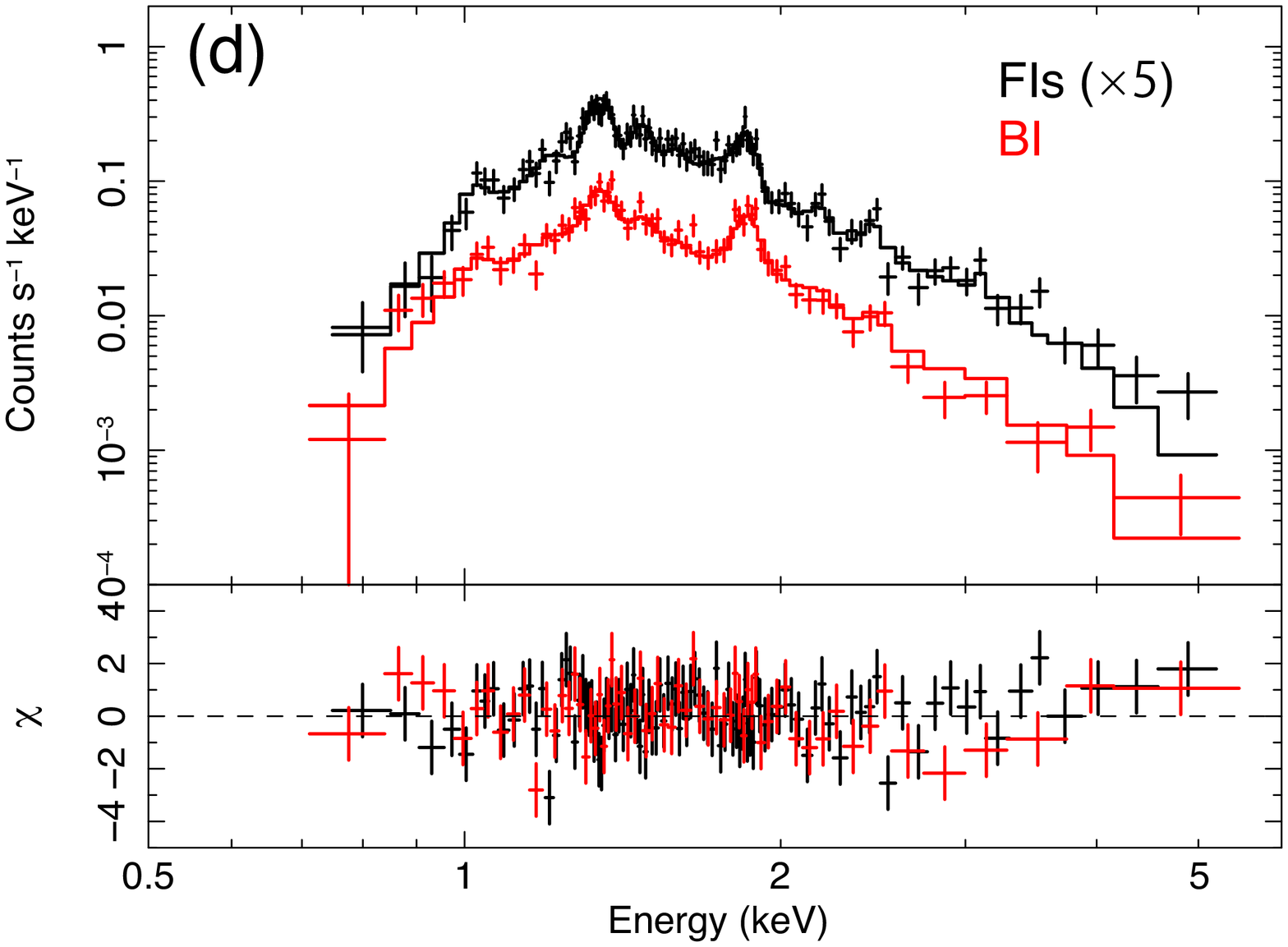}
  \end{center}
  \caption{Background-subtracted spectra with the best-fit models.
  Each panel shows the result of Reg~1 for the CIE (a) and the NEI (b) model and Reg~2 for the CIE (c) and  the NIE (d) model.
 Black and red data show the FI and BI spectra, respectively. For visibility, the FI spectrum is multiplied by 5. 
 The vertical error bar of each data point represents the 1$\sigma$ error. The Gaussian component for FI and BI are shown with green solid and dashed lines, respectively.}
\label{fig:reg1_spec}
\end{figure*}

There is no observation of a blank sky nearby W51C.
To estimate the background for the soft X-ray emission carefully, we sought a Suzaku archival dataset satisfying the following criteria: (i) 40$^{\circ}$ $\leq$ {\it l} $\leq$ 70$^{\circ}$, $-$0$\fdg$50 $\leq$
{\it b} $\leq $ $-$0$\fdg$20, (ii) exposure $>$ 40~ks, (iii) free from bright sources, and (iv) observed with normal clocking mode. 
As a result, we found that the background field (hereafter, BGD) listed in table~\ref{tab:obs_log} only meets the criteria.

The background emission of XIS consists of three components, i.e. NXB, cosmic X-ray background (CXB) and Galactic ridge X-ray emission (GRXE: e.g.,~\cite{Koyama89}). NXB is estimated by using {\tt xisnxbgen} as described in subsection~\ref{image_ana}. 
Since CXB has the uniform spectral shape and intensity in all sky~\citep{Kushino02}, we can model the  CXB spectrum in an arbitrary region.
GRXE has strong positional dependencies. 
The longitudinal and latitudinal intensity profiles of the Galactic center X-ray emission and GRXE Fe {\small {XXV}} K$\alpha$ (6.7~keV) line were modeled by~\citet{Uchiyama12}, which were obtained by the analysis of XIS data with $|l| < 125^{\circ}$ and $|b| < 5^{\circ}$. Using this model, we can derive the GRXE intensity ratio between two arbitrary regions.
Note that we did not use the result of~\citet{Kaneda97} which was obtained by the study of the latitudinal dependence of GRXE in the Scutum arm region (28$^\circ < l <~$29$^{\circ}$).
This is because they could not resolve the Fe\emissiontype{I}~K$\alpha$ (6.4~keV), XXV~K$\alpha$ and XXVI~Ly$\alpha$ (6.9~keV) lines and represented the spectrum of GRXE with one foreground emission (FE) plus one thermal plasma component, which is insufficient (see subsection~\ref{hard_bgd}).

\begin{table*}
  \caption{The best-fit results\footnotemark[$*$] of Reg~1 and Reg~2.}\label{tab:spec_result}
  \begin{center}
    \begin{tabular}{lcc}
      \hline
      \multicolumn{1}{l}{Parameter} & Reg~1 & Reg~2 \\\hline
      CIE &  &   \\
     $N_{\rm H}$ (10$^{22}$ cm$^{-2}$) & 1.56 (1.61) $_{-0.05}^{+0.06}$  &
	     2.01 (2.07) $_{-0.08}^{+0.09}$ \\
     $VEM$ (10$^{57}$ cm$^{-3}$)\footnotemark[$\dagger$] & 7.68 (7.74) $_{-0.96}^{+1.31}$  & 7.17 (7.47) $_{-1.01}^{+1.41}$ \\
     $T_{\rm e}$ (keV) & 0.49 (0.48) $_{-0.02}^{+0.02}$ & 0.56 (0.55) $_{-0.03}^{+0.02}$ \\
     $Z_{\rm Mg}$ (solar) & 1.91 (2.02) $_{-0.17}^{+0.18}$  & 1.53 (1.61) $_{-0.20}^{+0.20}$ \\
     $Z_{\rm Si}$ (solar) & 1.08 (1.14) $_{-0.12}^{+0.13}$ & 0.50 (0.51) $_{-0.09}^{+0.09}$ \\
     $Z_{\rm S}$ (solar) & 1.46 (1.49) $_{-0.29}^{+0.32}$ & 0.40 (0.40) $_{-0.18}^{+0.19}$ \\
     $\chi^2$/d.o.f & 448/348 &  200/160 \\\hline
     NEI & &  \\
     $N_{\rm H}$ (10$^{22}$ cm$^{-2}$) & 1.49 (1.51) $_{-0.05}^{+0.05}$ &
	    1.97 (2.02) $_{-0.07}^{+0.08}$ \\
     $VEM$ (10$^{57}$ cm$^{-3}$)\footnotemark[$\dagger$] & 4.14 (3.94) $_{-0.62}^{+0.70}$  & 5.06 (5.12) $_{-0.79}^{+1.01}$ \\
     $T_{\rm e}$ (keV) & 0.70 (0.69) $_{-0.05}^{+0.06}$  &  0.69 (0.68) $_{-0.05}^{+0.05}$ \\
     $n_{\rm e}t$ (10$^{11}$ cm$^{-3}$ s) & 1.77 (1.89) $_{-0.42}^{+0.57}$ &
	     2.97 (3.14) $_{-0.83}^{+1.15}$ \\
     $Z_{\rm Mg}$ (solar) & 1.73 (1.91) $_{-0.14}^{+0.16}$ &  1.40 (1.50) $_{-0.15}^{+0.16}$ \\
     $Z_{\rm Si}$ (solar) & 0.89 (0.97) $_{-0.10}^{+0.11}$ &  0.48 (0.50) $_{-0.08}^{+0.08}$ \\
     $Z_{\rm S}$ (solar) & 0.83 (0.88) $_{-0.19}^{+0.21}$ &  0.30 (0.30) $_{-0.14}^{+0.15}$ \\
     $\chi^2$/d.o.f & 383/347 &  168/159 \\
      \hline
     \multicolumn{3}{@{}l@{}}{\hbox to 0pt{\parbox{100mm}{
     \footnotemark[$*$] 
     The values in the parentheses are obtained by increasing the GRXE spectrum by a factor of 1.25, for systematics.\\
     \footnotemark[$\dagger$] Volume emission measure, $VEM = \int
     n_{\rm e}n_{\rm H}dV$, where $n_{\rm e}$ and $n_{\rm H}$ are the electron and
     Hydrogen densities, respectively, and $V$ is the emitting volume, assuming the distance of 6~kpc.}}}
    \end{tabular}
  \end{center}
\end{table*}

For the analysis of Reg~1 and Reg~2, we modeled the GRXE spectrum toward W51C by scaling that toward the BGD observation region. We first excluded the point sources and the region of $b\geq{-0}\fdg$20 from BGD because the analysis regions in W51C are located at $b<{-0}\fdg$20 (see figure~\ref{fig:cmap}).
The solid angle of the remaining region was 221~arcmin$^{2}$. 
Then we subtracted NXB and CXB to create the GRXE spectrum toward the BGD region.
We assumed the CXB spectrum to have a powerlaw with a photon index of 1.41 and an intensity of 5.4$\times$10$^{-15}$~erg~s$^{-1}$~arcmin$^{-2}$ in 2--10~keV band~\citep{Kushino02}. 
The absorption of interstellar medium (ISM:~\cite{Morrison83}) for CXB is attributed to Galactic H\emissiontype{I} medium of $\sim$0.8$\times$10$^{22}$~cm$^{-2}$~\citep{Kalberla05}\footnote{We used the H\emissiontype{I} calculator available at $\langle$http://heasarc.nasa.gov/cgi-bin/Tools/w3nh/w3nh.pl$\rangle$.} and molecular hydrogen of $\sim$0.2$\times$10$^{22}$~cm$^{-2}$~\citep{Dame01}. 
Then, the overall absorption column density for CXB is $N_{\rm H}^{\rm total}$~=~$N_{\rm H\emissiontype{I}}$\,+\,2$N_{\rm H_2}$~=~1.2$\times$10$^{22}$~cm$^{-2}$.
We assumed the solar abundance for the absorbing materials~\citep{Anders89}.

To derive a GRXE scaling factor of the W51C region against the BGD region, we used the GRXE intensity profile and then the scaling factors were obtained to be 1.76 and 1.86 for Reg~1 and Reg~2, respectively.
Since the profile of the GRXE intensity was estimated with an uncertainty of 25\%, we take into account it as the systematic error for the spectral analysis.

\subsubsection{Reg~1 and Reg~2}
\label{reg1_2}

To investigate the nature of the soft X-ray emission, we extracted the spectra from Reg~1 and Reg~2. 
For background subtraction, we first subtracted NXB and CXB considering the ISM absorption with the $N^{\rm total}_{\rm H}$ of 1.9$\times$10$^{22}$~cm$^{-2}$ and 2.8$\times$10$^{22}$~cm$^{-2}$ for Reg~1 and Reg~2, respectively.
These column densities are determined in the same manner as subsection~\ref{soft_bgd}. 
Then the scaled GRXE spectrum prepared in subsection~\ref{soft_bgd} was subtracted from each region after correcting the effective area.
The background-subtracted spectra are shown in figure~\ref{fig:reg1_spec}.
We can see K$\alpha$ emission lines from He-like (hereafter, He$\alpha$)
neon (Ne), magnesium (Mg), silicon (Si) and sulfur (S).

For Reg~1, we first tried to fit the spectrum with an optically thin thermal plasma in CIE with an ISM absorption.
The abundances of the noticeable elements Mg, Si and S ($Z_{\rm Mg}$, $Z_{\rm Si}$ and $Z_{\rm S}$) were free parameters. 
Those of Ne and Fe were fixed at the solar value because the Ne-K and Fe-L lines are overlapping and the heavy absorption make it hard to determine them independently. 
The other elements were also fixed at the solar values.
This model failed to reproduce the spectrum with a $\chi^{2}$/d.o.f of 593/349, leaving large residuals around 0.9~keV, 1.2~keV, and $>$ 3.0~keV.
The first residual around 0.9~keV corresponds to the Ne He$\alpha$.
The second is probably due to the lack of a number of Fe L-shell lines emitted at the transitions from highly excited states of n $\geq$ 5~\citep{Brickhouse00} in the plasma model. Thus, a Gaussian model was artificially added at 1.2~keV. The fit was largely improved to a $\chi^{2}$/d.o.f of 448/348
although the residuals around 0.9~keV and $>$ 3~keV still remain. The best-fit spectra and parameters for the one CIE model are given in figure~\ref{fig:reg1_spec}a and table~\ref{tab:spec_result},
respectively. 
The value of $Z_{\rm Mg}$ was obtained to be 1.91$^{+0.18}_{-0.17}$ solar, which is significantly
higher than the solar value.

We next examined a non-equilibrium ionization (NEI) model.
In the NEI model, the ionization parameter $n_{\rm e}t$ was an additional free parameter, where $n_{\rm e}$ and $t$ are the electron density and the elapsed time after the shock heating.
The fit was significantly improved to a $\chi^{2}$/d.o.f of 383/347 without a notable residual around 0.9~keV (figure~\ref{fig:reg1_spec}b and table~\ref{tab:spec_result}).
The values of $T_{\rm e}$ and $n_{\rm e}t$ are obtained to be
0.70$^{+0.06}_{-0.05}$~keV and 1.77$^{+0.57}_{-0.42}~\times $10$^{11}$~cm$^{-3}$ s, respectively.
The obtained $Z_{\rm Mg}$ of 1.73$^{+0.16}_{-0.14}$ is still significantly higher than the solar value.

We also checked whether the plasma is over-ionized since the excess around 0.9~keV and $>$ 3~keV could be caused by the radiative recombination continua (RRCs).
RRCs appear when the recombination process dominates over the ionization one in the over-ionized plasma ($T_{\rm z}\,>\,T_{\rm e}$).
If RRCs generate the notable residuals, the excesses around 0.9~keV and $>$~3~keV would correspond to the K-shell recombinations of H-like O and Si ions, respectively.
Hence, we set $Z_{\rm O}$ and $T_{\rm z}$ to be free in addition to the above CIE model.
As a result, we obtained $T_{\rm z}$/$T_{\rm e}$ of 0.83$_{-0.11}^{+0.14}$, which is rather consistent with an NEI (under-ionized) picture than an over-ionized one.

For Reg~2, we examined the spectrum with a CIE model affected by
the absorption of ISM. 
This model reproduced the spectrum with a $\chi^{2}$/d.o.f of 200/160.
The obtained spectrum and parameters are shown in figure~\ref{fig:reg1_spec}c and
table~\ref{tab:spec_result}. We also tried an NEI model. 
The results were significantly improved to a $\chi^{2}$/d.o.f of 168/159, and the normalization of the Gaussian at 1.2~keV became 0. The best-fit spectra and the parameters are shown in figure~\ref{fig:reg1_spec}d and table~\ref{tab:spec_result}.
We also checked the probability of over-ionization with the same procedure as used for Reg~1.
As a result,  $T_{\rm z}$/$T_{\rm e}$ was obtained to be 0.79$_{-0.09}^{+0.10}$ and then the plasma of Reg~2 is not over-ionized.
Thus, we again adopted the NEI model as the best-fit for Reg~2. 
The plasma properties are essentially the same as Reg~1.

The obtained results of Reg~1 and Reg~2 are inconsistent with that
obtained by Chandra~\citep{Koo05}. 
They reported that the spectra were represented by a CIE plasma with temperature of 0.3--0.5~keV and enhanced $Z_{\rm S}$. 
This might be caused by the inappropriate background estimation without considering the latitudinal dependence of GRXE in the Chandra data analysis.

\subsection{Hard X-Ray Emission}
\label{hard_spec}

\subsubsection{Background Modeling}
\label{hard_bgd}

The hard X-ray emission widely extends around molecular clouds associated with W51B but the distance to the emitting region is unclear.
We can estimate the distance from $N_{\rm H}$.
Thus, the accurate estimation of the background emission toward Reg~3 is required.
We first modeled the spectrum of the BGD region and then modified it to Reg~3, considering the difference of the Galactic absorption column densities and the intensities of GRXE.
The modeling method is described below.

\begin{figure}
  \begin{center}
    \FigureFile(80mm,30mm){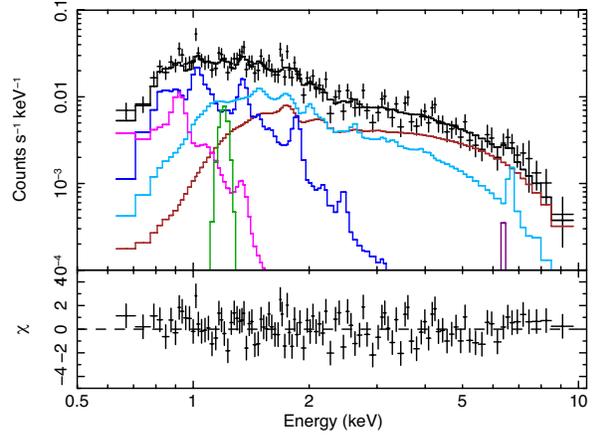}
  \end{center}
  \caption{XIS spectrum of the BGD region with the best-fit models in equation
 (\ref{bgd_model}).
For visibility, data and models for only the FI are displayed.
Black data shows the XIS data while a brown curve shows the CXB component.
Magenta, blue, cyan, green, and purple curves show the best-fit models of CIE1, CIE2, CIE3, Gaus1 and Gaus2 components in equation (\ref{bgd_model}), respectively.}
\label{fig:bkg_spec}
\end{figure}

\begin{table*}
  \caption{The best-fit result of the BGD region.}\label{tab:bgd_result}
  \begin{center}
    \begin{tabular}{lccc}
      \hline
            \multicolumn{1}{l}{Parameter} & Foreground & Medium & High \\
      \multicolumn{1}{l}{} & (CIE1) & (CIE2) & (CIE3) \\\hline
     $N_{\rm H}$ (10$^{22}$ cm$^{-2}$) & 0.49 $_{-0.12}^{+0.25}$ & \multicolumn{2}{c}{0.58 (fixed)}\\
     $T_{\rm e}$ (keV) & 0.11 $_{-0.03}^{+0.06}$ & 0.49 $_{-0.13}^{+0.09}$ & 3.82 $_{-0.86}^{+1.49}$  \\
     $Z_{\rm Fe}$ (solar) & 1.00 (fixed) & \multicolumn{2}{c}{0.61 $_{-0.19}^{+0.23}$}  \\
      $f_{0.5-10}$\footnotemark[$*$]  & 1.60 (52.9) & 2.25 (11.7) & 7.25 (10.6) \\\hline
      $f_{6.4}$\footnotemark[$\dagger$]  & \multicolumn{3}{c}{6.16 $_{-6.16}^{+22.7}$}  \\\hline
     $\chi^2$/d.o.f & \multicolumn{3}{c}{186/167} \\
      \hline
     \multicolumn{4}{@{}l@{}}{\hbox to 0pt{\parbox{100mm}{
     \footnotemark[$*$] Observed flux in the 0.5--10~keV band in units of 10$^{-13}$~erg~s$^{-1}$~cm$^{-2}$. Values in parentheses are the absorption-corrected values.\\
     \footnotemark[$\dagger$] Observed photon flux of 6.4~keV line in units of  10$^{-7}$~photons~s$^{-1}$~cm$^{-2}$.}}}
    \end{tabular}
  \end{center}
\end{table*}

As mentioned in subsection~\ref{soft_bgd}, the emission toward the BGD region consists of NXB, CXB, and GRXE.
Unlike subsection~\ref{soft_bgd}, we here modeled the GRXE spectrum by a three temperature CIE model~\citep{Uchiyama09}.
A Gaussian was added at 1.2~keV to compensate the lack of the Fe L-shell lines (see subsection~\ref{reg1_2}).
We also considered the Fe {\small {I}} K$\alpha$ line of GRXE~\citep{Ebisawa08} to represent the emission in the hard X-ray band more accurately. Because the lowest temperature component of GRXE is considered as a foreground emission, the absorption for it was independent of that for the other two CIE components.
After subtracting NXB, we fitted the spectrum of BGD, as shown in figure~\ref{fig:bkg_spec}, using the following spectral model,
\begin{eqnarray}
\label{bgd_model}
\nonumber
{\rm Absm1}&{\times}&{\rm Powerlaw1}+{\rm Absm2{\times}CIE1}\\
&+&{\rm Absm3}{\times}{\rm (CIE2+CIE3}+{\rm Gaus1+Gaus2)}.
\end{eqnarray}
Powerlaw1, CIE1, CIE2, and CIE3 represent CXB, FE, the medium and high-temperature component of GRXE. Gaus1 and Gaus2 are a Gaussian at 1.2~keV and 6.4~keV lines, respectively.
Absm1, Absm2, and Absm3 represent the ISM absorptions for CXB, FE, and medium+high-temperature components of GRXE, respectively.
The spectral parameters for CXB including Absm1 were fixed at those used in subsection~\ref{soft_bgd}. 
GRXE has an axial symmetric structure for the Galactic center and its longitudinal scale length of 45$^\circ$~\citep{Uchiyama12} is smaller than that of absorbing gas of $\sim$100$^\circ$~\citep{Amores05}.
Therefore, we assumed that the GRXE emitting plasma is located at half distance to the edge of the Galaxy along the line of sight and the gas is divided to near and far side components for the plasma.
Hence, we set Absm3~=~1/2$\times$Absm1. For FE, we treated Absm2 as a free parameter.
The elemental abundances of CIE1 were fixed at the solar value. $Z_{\rm Fe}$ of CIE2 and CIE3 were set to be common and the other abundances were fixed at the solar value.
The fitting model with equation~(\ref{bgd_model}) well represented the spectrum with a $\chi^{2}$/d.o.f of 186/167 (see figure~\ref{fig:bkg_spec}), where the best-fit parameters are listed in table~\ref{tab:bgd_result}.

To derive the background emission in Reg~3, we scaled the normalization of CIEs and Gaussians of the BGD region by a factor of 1.93, which was obtained by the same estimation as used in subsection~\ref{soft_bgd}.
For Absm1 and Absm3, we adopted the column densities of throughout the Galaxy and half of it toward Reg~3, respectively (3.3$\times$10$^{22}$~cm$^{-2}$ and 1.7$\times$10$^{22}$~cm$^{-2}$).
The $N_{\rm H}$ of Absm2 and powerlaw model for CXB were assumed to be the same as those of BGD.
Then, we obtained the background model for Reg~3.

\subsubsection{Reg~3}
\label{reg3}
 
To investigate the nature of  the extended hard X-ray emission, we analyzed Reg~3 as shown in figure~\ref{fig:cmap}b.
The bright hard X-ray sources, the PWN candidate, and two H\emissiontype{II} regions G49.0$-$0.3 and G49.2$-$0.3 were excluded from the region.

\begin{figure}
  \begin{center}
   \FigureFile(80mm,30mm){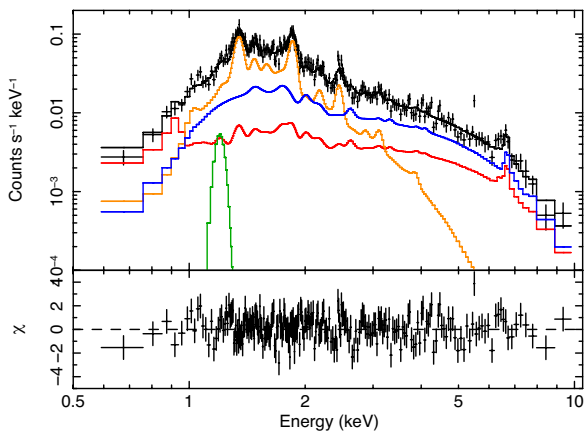}
  \end{center}
  \caption{XIS spectrum of Reg~3 with the best-fit model of equation (\ref{spec_model1}). For visibility, data and models for only the FI are displayed. Black data shows the XIS data. 
 Red, orange, green, and blue curves show the models for Background model, NEI, Gaus3 and CIE4 components in equation (\ref{spec_model1}), respectively.}
\label{fig:spec_reg3}
\end{figure}

\begin{table*}
  \caption{The fit results of Reg~3\footnotemark[$*$].}\label{tab:reg3_result}
  \begin{center}
    \begin{tabular}{lcc}
      \hline
      \multicolumn{1}{l}{Parameter} & Model of equation (\ref{spec_model1})
     & Model of equation (\ref{spec_model2})\\\hline
     NEI\footnotemark[$\dagger$] &  &  \\
     $N_{\rm H}$ (10$^{22}$ cm$^{-2}$) & 2.42 (2.41) $_{-0.10}^{+0.10}$ &  2.40 (2.38) $_{-0.12}^{+0.12}$ \\
     $VEM$ (10$^{57}$ cm$^{-3}$)\footnotemark[$\ddagger$] & 
	 5.02 (5.05) $_{-0.43}^{+0.45}$  &  4.66 (4.86) $_{-0.52}^{+0.54}$  \\\hline
     Thermal (CIE4) & &   \\
     $N_{\rm H}$ (10$^{22}$ cm$^{-2}$) & 0.86 (0.98) $_{-0.13}^{+0.17}$ & $\cdots$ \\
     $VEM$ (10$^{58}$ cm$^{-3}$)\footnotemark[$\ddagger$] &
	 6.79 (6.50) $_{-0.75}^{+0.86}$  & $\cdots$ \\
     $T_{\rm e}$ (keV) & 5.08 (4.93) $_{-0.83}^{+1.12}$ & $\cdots$ \\
     $Z_{\rm Fe}$ (solar) & 0.24 (0.23) $_{-0.16}^{+0.16}$ & $\cdots$ \\
     Luminosity\footnotemark[$\S$] & 6.53 (6.14) $_{-0.72}^{+0.83}$  & $\cdots$ \\\hline
     Nonthermal (Powerlaw2) & & \\
     $N_{\rm H}$ (10$^{22}$ cm$^{-2}$) & $\cdots$ & 1.15 (1.30) $_{-0.19}^{+0.25}$ \\
     Photon index & $\cdots$ & 2.23 (2.29) $_{-0.18}^{+0.19}$ \\
     Luminosity\footnotemark[$\S$] & $\cdots$  & 7.02 (6.63) $_{-1.76}^{+2.41}$ \\\hline
     $\chi^{2}$/d.o.f & 443/375 & 445/376 \\\hline
     \multicolumn{3}{@{}l@{}}{\hbox to 0pt {\parbox{118mm}{
     \footnotemark[$*$] The values in the parentheses are obtained by increasing the GRXE spectrum by a factor of 1.25, for systematics.\\
     \footnotemark[$\dagger$] The parameters except for $VEM$ are fixed at the values of the NEI model for Reg~1.\\
     \footnotemark[$\ddagger$] 
     Volume emission measure, $VEM = \int
     n_{\rm e}n_{\rm H}dV$, where $n_{\rm e}$ and $n_{\rm H}$ are the electron and
     Hydrogen densities, respectively, and $V$ is the emitting volume, assuming the distance of 6~kpc.\\
     \footnotemark[$\S$] Unabsorbed luminosity in the 2.0--10.0~keV band in
     units of 10$^{33}$ erg s$^{-1}$, for the distance of 6~kpc.}}}
    \end{tabular}
  \end{center}
\end{table*}

We show the NXB-subtracted spectrum of Reg~3 in figure~\ref{fig:spec_reg3}.
In this region, the soft thermal X-rays from W51C also contribute as background emission.
We represented the W51C emission by the NEI model of Reg~1 (see subsection~\ref{reg1_2}).
We did not use the Reg~2 results, since the errors are larger but the results are consistent with those of Reg~1.
We tried to fit the hard X-ray emission with a CIE or a powerlaw model as follows,
\begin{eqnarray}
\label{spec_model1}
\nonumber
{\rm Background~model}&+&{\rm Absm4}{\times}({\rm NEI+Gaus3})\\
&+&{\rm Absm5}{\times}{\rm CIE4},
\end{eqnarray}
or
\begin{eqnarray}
\label{spec_model2}
\nonumber
{\rm Background~model}&+&{\rm Absm4}{\times}({\rm NEI+Gaus3})\\
&+&{\rm Absm5}{\times}{\rm Powerlaw2}.
\end{eqnarray}
Here, the background model consists of CXB and the scaled GRXE.
NEI and Gaus3 are the W51C emission and the 1.2~keV line making up for the lack of Fe-L shell lines, respectively.
We set the normalization of the NEI and Gaus3 to be free. The $N_{\rm H}$ of Absm4 and Absm5, the normalization, $T_{\rm e}$, and $Z_{\rm Fe}$ of CIE4, and the normalization and photon index of Powerlaw2 were also free parameters.
The fitted spectrum and parameters are shown in figure~\ref{fig:spec_reg3} and table~\ref{tab:reg3_result}.

 The obtained value of $N_{\rm H}$ of Absm5 is smaller than $\sim$3$\times$10$^{22}$~cm$^{-2}$, the total column density toward this direction, suggesting that the hard X-rays is Galactic origin.
This value is similar to $\sim$1.5$\times$10$^{22}$~cm$^{-2}$ derived by the X-ray spectral analysis of H\emissiontype{II} regions, G48.9$-$0.3 and G49.0$-$0.3, located at the front side of the clouds~\citep{Koo02}.
Therefore, we consider that the hard X-ray emitting region is located around the clouds and W51C, although this is not conclusive due to the systematic uncertainties stemming from the estimation of GRXE and spectral modeling of SNR emission.

The analyzed spectrum of Reg~3 contains emission from point sources associated with W51B and background sources. We estimated the contribution of these point sources to the extended emission as described below.

First, we investigated a time variability of the X-ray emission from point sources.
The image of FIs in the 2.5--8.0~keV band was compared with the Chandra image convolved with the PSF of the Suzaku XRT. 
We found a compact source, whose flux is similar to that of G49.0$-$0.3, in the convolved Chandra image, located at ($l$, $b$)~=~(49$\fdg$0206, $-$0$\fdg$2629), while there is apparently no object in the FIs.
It might be a compact source such as an active galactic nucleus like a blazar or a flare star which have a  short term variability. We did not find other differences between the two images.

Next, we estimated the total X-ray luminosity of the point sources in Reg~3 using the Chandra data.
The total photon counts of the point sources in Reg~3 without the above bright source were converted to fluxes in the 2.0--10~keV band by using {\tt WebPIMMS}\footnote{http://heasarc.gsfc.nasa.gov/Tools/w3pimms.html} assuming the emission from a CIE plasma attenuated by absorption of ISM. 
We assumed 3~keV and the solar values for the temperature and abundances of the CIE model, respectively, considering an emission model for typical OB-type stars and young stellar objects~\citep{Schulz01}.
Because the distances to the sources are unknown, we adopted the $N_{\rm H}$ of 2.1$\times$10$^{22}$~cm$^{-2}$, which is an intermediate value obtained by the X-ray analysis toward W51B~\citep{Koo02}.
Using the obtained unabsorbed flux, the net luminosities of the point sources are 1.02$\times$10$^{33}$~erg~s$^{-1}$.

We cannot distinguish which model (thermal or nonthermal) is statistically plausible for the present data.
The total flux of the point sources toward Reg~3 has only minor contribution to that of the hard X-rays using the background model, even considering the systematic error for the intensity profile of GRXE (see subsection~\ref{soft_bgd}).

\section{Discussion}
\label{discussion}

\subsection{Contribution of Suprathermal Electrons to Ionization of Thermal Plasma}
\label{RRC}


In SNRs, electrons injected into the first order Fermi acceleration are expected to have suprathermal energies. 
Because such electrons can effectively ionize ions in thermal plasma, we can consider the possibility  that the electrons make the plasma over-ionized (e.g.,~\cite{Kato92}).
Cosmic rays responsible for gamma-ray emission in SNRs are accelerated around the shock whereas X-ray-emitting plasmas are formed in the downstream of the shock.
Thus the cosmic rays and the plasmas would interact with each other.
Note that we cannot resolve two regions due to the lack of angular resolution of a current gamma-ray telescope. The GeV emitting region in W51C overlaps with Reg~1 and Reg~2~\citep{W51C}.
However, we could not find any significant features of the over-ionized plasma in this SNR (subsection~\ref{reg1_2}). 
One possibility is that the suprathermal electrons exist but the elapsed time of the ionization is too short to make the plasma over-ionized. 
To examine this idea, we here estimate the elapsed and characteristic timescales of ionization by suprathermal electrons.

We assume the energy distribution of the suprathermal electrons to be quasi-thermal, which means it is roughly a Maxwellian. 
We also assume the $``$temperature$"$ of the electrons to be 10~keV. 
In this case, the characteristic timescale of ionization by suprathermal electrons can be represented by $\langle\sigma{v}\rangle^{-1}$, where $\sigma$ and $v$ are the cross section of collisional ionization and the velocity of suprathermal electrons, respectively. We consider the ionization of H-like Mg ions.
By retrieving the value of $\sigma$ from the {\tt ALADDIN} database\footnote{$\langle$http://dpc.nifs.ac.jp/aladdin$\rangle$}, $\langle\sigma{v}\rangle^{-1}$ for the above situation is derived to be $\sim$5$\times$10$^{10}$~cm$^{-3}$~s.

The ionization parameter of suprathermal electrons is $n'_{\rm e}t'$, where $n'_{\rm e}$ and $t'$ are the number density of suprathermal electrons and the elapsed time of the ionization. 
Because $n'_{\rm e}$ is unknown, we assume that $n'_{\rm e}$ scales $n_{\rm e}$, the number density of thermal electrons. 
We take the density ratio, $n'_{\rm e}$/$n_{\rm e}$, to be 10$^{-3}$ obtained around the shock front of SN~1006 where efficient acceleration occurs.
The ratio should be taken as the upper limit for the middle-aged SNR W51C since \citet{Bamba03} derived the amount of nonthermal electrons including suprathermal ones which have the energy of 10--100~keV.

The value of the ionization parameter $n_{\rm e}t$ for the thermal electrons is about 2$\times$10$^{11}$~cm$^{-3}$~s from the spectral analysis~(subsection~\ref{reg1_2}). 
Then, by scaling this value for suprathermal electrons with the above $n'_{\rm e}$/$n_{\rm e}$, we derive $n'_{\rm e}t'$ to be 2$\times$10$^{8}$~cm$^{-3}$~s under the assumption of $t'\sim{t}$, where $t$ is the elapsed time of ionization by the thermal ones.
The value of $n_{\rm e}'t'$ is much shorter than the characteristic timescale of the ionization by suprathermal electrons.
If the suprathermals are more abundant due to the second order Fermi acceleration processes (e.g., \cite{Bykov00}), $n'_{\rm e}$/$n_{\rm e}$ is about a few\,\%~\citep{Masai02}.
Even in this case, the value of $n_{\rm e}'t'$ is still by one order of magnitude shorter than the characteristic one. 
Hence, the ionization by suprathermal electrons is negligible and the non-detection of an over-ionized plasma in W51C do not rule out the possibility that the cosmic-ray acceleration is ongoing there.
Note that for a more realistic discussion, the time evolution of $n_{\rm e}$ should be taken into account.

\subsection{Origin of Hard X-Ray Emission}
\label{dis:hard_xray}

The hard X-ray emission in Reg~3 is represented by the thermal plasma model with $\sim$5~keV or the nonthermal emission with the photon index of $\sim$2.2. The obtained $N_{\rm H}$ is quite smaller than that throughout the Galaxy, suggesting a Galactic origin. 
The emission probably have diffuse nature because the luminosity of 1$\times$10$^{34}$~erg~s$^{-1}$ in the 0.5--10 keV band cannot be explained by the ensemble of the point sources located in the region.
A PWN is unlikely to be the origin because no pulsars have been found in Reg~3.
Based on these results, we discuss the possible origins of the diffuse hard X-ray emission.

\subsubsection{Stellar Winds from OB stars in W51B}
\label{W51B_origin}

The hard X-ray emission overlaps the molecular clouds associated with the star-forming region W51B. There are at least four H\emissiontype{II} regions around Reg~3. Such an environment leads us to the idea that the hard X-rays originate from shock-induced bubbles due to stellar winds from OB stars (e.g.,~\cite{Weaver77}). Stellar wind bubbles would have thermal and/or nonthermal emission (e.g.,~\cite{Wolk02,Ezoe06}).

We first consider the possibility that the hard X-ray photons originate
from shock-heated thermal plasma via the stellar winds.
The volume of the emission region is approximated to be a sphere of 10-pc radius estimated from the apparent size of Reg~3 and the line of sight depth of the clouds~\citep{Carpenter98}.
From the equation (3) of \citet{Ezoe06}, the total thermal energy with the temperature of 5 keV is obtained to be $\sim$2$\times$10$^{50}$~erg.
As the age of the H\emissiontype{II} regions around Reg~3 is about 1~Myr~\citep{Kim07}, the required power is $\sim$6$\times$10$^{36}$~erg~s$^{-1}$.
%
%
%
%
According to~\citet{Kim07}, the spectral types of the most massive stars in H\emissiontype{II} regions G49.0$-$0.3 and G49.2$-$0.3 are O9 and O4, respectively. 
\citet{Muijres12} estimated that the mass loss rate and the velocity of the stellar winds of an O4 star is about 10$^{-5.5}$~M$\odot$~yr$^{-1}$ and 4000~km~s$^{-1}$, respectively. Then the kinetic power of the stellar winds is calculated to be $\sim$10$^{37}$~erg~s$^{-1}$ from the equation (6) of \citet{Ezoe06}.
Therefore, thermal emission from stellar winds might be the origin of the hard X-rays.

Next, we discuss nonthermal origin with stellar winds; bremsstrahlung from nonthermal ($>$~10~keV) electrons, inverse Compton (IC) scattering by hundreds-of-MeV electrons, and synchrotron emission from TeV electrons. We can rule out the bremsstrahlung emission on the basis of the spectral index. 
In a dense environment (the gas density $>$~10~cm$^{-3}$), bremsstrahlung emission dominates the hard X-ray band. The spectral index of energetic electrons should be flatter than the initial spectrum because lower energy electrons lose their energy more quickly than more energetic ones via the Coulomb loss. The resulting X-ray emission has a spectral index of $\sim$ 1.2~\citep{Uchiyama02}, but the observed index is $\sim$2.2.

The possibility of IC scattering can also be ruled out.
Electrons with energies of several hundred MeV (with a Lorentz factor of 10$^2$--10$^3$) can produce hard X-ray emission by scattering the IR emission from star lights in W51B.
The steep hard X-ray spectrum could be the result of synchrotron loss.
However, the cooling time of the electrons even with the magnetic field of
100~$\mu$G is $\sim$10~Myr, which is much longer than the age of W51B.

Synchrotron emission from tens of TeV electrons produced by stellar winds could explain the diffuse X-rays.
We roughly estimated the amount of the energy of the TeV electrons responsible for the X-ray emission in the 0.5--10~keV band assuming a magnetic field of 10~$\mu$G and then it is obtained to be about 1$\times$10$^{45}$~erg. 
Since the cooling time of the TeV electrons is a few 10$^{3}$~yr, the energy should be injected within the timescale. Then, the average energy input is constrained to be larger than about 10$^{34}$~erg~s$^{-1}$, which can be supplied by stellar winds from the OB stars around W51B.




\subsubsection{SNR W51C}
\label{SNR_origin}

The hard diffuse X-ray emission also overlaps the western part of the shell of W51C.
Thermal emission from the SNR is less likely to be the origin of the diffuse X-rays since the typical plasma temperature of SNRs is below a few keV.
Therefore, nonthermal origin is more preferable interpretation.
We can immediately rule out the possibility of bremsstrahlung and IC scattering because the spectral index of the X-rays is soft ($\sim$2.2: subsection~\ref{W51B_origin}).

If particle acceleration terminates at the early Sedov stage, synchrotron emission from tens of TeV electrons cannot explain the hard X-rays since the cooling time of the high energy electrons is a few 10$^{3}$~yr under a magnetic field of $\sim$10~$\mu$G.
Alternatively, the observed X-ray flux might be explained by the synchrotron emission from secondary particles produced by decay of charged pions generated by interaction of cosmic rays in dense gas.
However, the synchrotron X-ray emission from secondary particles is fairly dim in the hard X-ray band even with higher gas density than 100~cm$^{-3}$ in an SNR having a similar $\gamma$-ray spectrum and an age to W51C (e.g.~\cite{G8.7}).


On the other hand, if particle acceleration takes place over the SNR age, tens-of-TeV electrons can exist in middle-aged SNRs~\citep{Sturner97,Nakamura12} and then synchrotron emission from the TeV electrons can explain the X-ray spectrum.
The detection of possible synchrotron X-rays in the middle-aged SNR W44 might be related to TeV electrons~\citep{Uchida12} and supports the above scenario.
Assuming the acceleration over the SNR age, very efficient acceleration is unlikely to operate because $n'_{\rm e}t'$ exceeds the characteristic timescale of the ionization by suprathermal electrons under a efficient acceleration (see subsection \ref{RRC}).
For the acceleration only in the late phase of the SNR, the particle acceleration mechanism might be related to the interaction of the SNR shells with molecular clouds. \citet{Inoue09} demonstrated that shock-compressed shells become turbulent owing to the inhomogeneity of preshock density and then amplification of magnetic field takes place in the shell. Particles could be accelerated at the secondary shocks arising from the turbulent flows.
However, the secondary shocks with a velocity of hundreds~km~s$^{-1}$ cannot accelerate electrons up to TeV energies due to fast synchrotron cooling.


An alternative idea is that high energy electrons are accelerated by SNR(s) in tenuous ISM, where the SNRs would effectively accelerate particles without deceleration of the shocks~(e.g.,~\cite{Tang05}).
Such an environment might be provided by core-collapse supernova explosions before the birth of W51C.
This idea is supported by the enhanced Mg abundance in Reg~1 and Reg~2. Because the main component of the plasma is possibly ISM (\cite{Koo02} and this work; see $Z_{\rm Si}$ and $Z_{\rm S}$ in table~\ref{tab:spec_result}), the enhanced Mg would come from the past explosion of 20--25~M$\odot$ stars, which supply a large amount of Ne and Mg~\citep{Tsujimoto95}.
The existence of two PWN candidates around W51C provides another clue for the past core-collapse supernovae in this region.

\section{Conclusion}
\label{conclusion}

We investigated the X-ray emission toward the middle-aged SNR W51C and its environment with
XIS onboard Suzaku.
The soft X-ray emission from the plasma of W51C is well represented by an optically thin thermal plasma in NEI with the temperature of $\sim$0.7~keV with the enhanced  Mg abundance.
Although the thermal emission region overlaps with the GeV emitting region where cosmic rays are accelerated in SNRs, there is no hint of an over-ionized plasma.
These results are consistent with the situation that the elapsed time of the ionization by suprathermal electrons is too short to make the plasma over-ionized.

The hard X-ray emission is spatially coincident with the molecular
clouds associated with W51B.
The spectrum of Reg~3 is represented by an optically thin thermal plasma with a temperature of $\sim$5~keV or a powerlaw model with a photon index of $\sim$2.2.
The emission have diffuse nature because the luminosity of 1$\times$10$^{34}$~erg~s$^{-1}$ in the 0.5--10~keV band cannot be explained by the ensemble of point sources. 
The smaller $N_{\rm H}$ than that throughout the Galaxy suggests a Galactic origin but a PWN is unlikely the candidate because no pulsars have been found in Reg~3.
The origin of the hard X-rays can be interpreted with the stellar wind scenarios.
The X-ray emission can be also interpreted by the synchrotron X-rays due to the TeV  electrons continuously accelerated by the SNR W51C after the free expansion phase or in tenuous ISM while it cannot be explained by the synchrotron emission either of leptonic or hadronic process of particles accelerated by the early Sedov phase.

%
%
%
%
%

\bigskip

The authors thank the anonymous referee for helpful comments.
We also thank Ryo Yamazaki, Kuniaki Masai, Yuichiro Ezoe, and Shigeo Yamauchi for valuable discussions and comments. We also thank Bon-Chul Koo for the provision of the radio image and the H\emissiontype{I} map, Hyosun Kim for the estimation of the number of OB stars, and Hideki Uchiyama for the support on the estimation of GRXE.  
This work was supported in part by the Grant-in-Aid for Scientific Research of the Ministry of Education, Culture, Sports, Science and Technology of Japan; 22740167 (H. Katagiri), 24840036~(M. Sawada), 22684012~(A. Bamba) and 60272457~(Y. Fukazawa).

\end{document}